\documentclass[12pt]{article}
\newcommand{\LI}{\hbox to\hsize}

\newcommand{\BEQ}{\begin{equation}}
\newcommand{\EEQ}{\end{equation}}

\newcommand{\DEG}[1]{\mbox{$ #1^{\rm o}$}}
\newcommand{\gappr}{\mbox{$\stackrel{>}{\sim}$}} % greater than, approximately
\newcommand{\incircle}[1]{\mbox{{\hbox{$\bigcirc$}\kern-0.7em
\lower0.05ex\hbox{\mbox{{\scriptsize\rm #1}}}}}}
\newcommand{\VIZ}{\mbox{\em viz.\/ }}
\newcommand{\CF}{\mbox{\em cf.\/ }}
\newcommand{\IE}{\mbox{\em i.e. \/}}
\newcommand{\ETAL}{\mbox{\em et. al.\/ }}
\newcommand{\EG}{\mbox{\em e.g.\/ }}
\newcommand{\tipacdate}[2]{\begin{flushright}
JHU--TIPAC--#1 \\
#2
\end{flushright}
\vspace*{8mm}}
\newcommand{\Title}[1]{\begin{center}
{\Large\bf #1}
\end{center}}
\newcommand{\MI}{\vspace*{5mm} \begin{center}
G. Domokos and S. Kovesi--Domokos\\[2mm]
The Henry A. Rowland Department of Physics and Astronomy\\
The Johns Hopkins University\footnote{E--mail:~skd@haar.pha.jhu.edu}\\
Baltimore, MD 21218
\end{center}}
\newcommand{\Abstract}[1]{\vspace*{4mm} \begin{center} \begin{quote}
#1 \end{quote} \end{center}\vspace*{4mm}}
\newcommand{\centerfigure}[2]{\leavevmode
 \begin{center}
 \epsfbox[108 160 1230 530]{#1}
\vspace{1truemm}
#2
\end{center}}
\begin{document}
\input epsf
\tipacdate{98004}{April 1998}
\Title{Observation of Ultrahigh Energy Neutrino Interactions by
Orbiting Detectors}
\MI
\Abstract{Orbiting detectors will be able to observe showers initiated 
by neutrinos penetrating the Earth and interacting close to their
exit point.
There is a correlation between the impact parameter of 
the incident neutrino and its energy. We study the development of upward
going, neutrino induced showers in the atmosphere.}
\section{Introduction}
\label{sec:intro}
The study of extraterrestrial, ultrahigh energy neutrinos is
important from at least two points of view. First, they probably
carry astrophysical information about point sources which are optically
thick at (almost) all wavelengths, for instance, an active galactic
nucleus (AGN). Second,  incoming neutrinos of 
laboratory energies of the order of $10^{18}$eV (or even higher)
are expected to be emitted by the highest energy sources.  In a collision
with a nucleon, the CMS energy is, therefore, of the order of
40~TeV, far exceeding the CMS energies available in accelerator generated
neutrino beams. Although such events are expected to be
rare ones,  they provide a unique window on physics beyond
the Standard Model.
For both reasons  it is important to refine the observation of 
ultrahigh energy (UHE) neutrinos. 
The main purpose of this paper is to sketch a novel way of observing
such interactions.

The basic principle is a very simple one. Neutrino--nucleon interactions have
a cross section which is a well known, monotonically increasing function
of the incident neutrino energy; for a recent calculation, up to neutrino
energies of $10^{12}$GeV,
\CF ref.~\cite{gandhi}. Consequently, {\em on the average\/}, a neutrino
penetrating the Earth will interact after a well defined distance
beneath the surface. If the interaction occurs  reasonably close to the exit
point of the neutrino, the ensuing shower develops largely in the
atmosphere and it can be detected by means of an orbiting detector, 
such as OWL
(Orbiting Wide angle Light collector) or AIR WATCH. Such detectors 
are suitable
for a study of the longitudinal development of  showers. Thus, a shower 
developing upwards gives a unique signature of a neutrino interaction. 
Further,
one obtains some information about the primary energy of the incident
neutrino: the Earth acts as an energy filter. Indeed, given the impact 
parameter of the incident neutrino, its (energy dependent) interaction
mean free path has to be approximately equal to the length of the chord
of the trajectory inside the Earth. If it is considerably shorter
(high energies), the interaction and the development of the shower
takes place largely inside the Earth: no atmospheric shower is observed.
Conversely, at low energies, the neutrino penetrates the Earth without
interacting. Once it leaves the Earth, its chance of interacting within 
the atmosphere is rather slim. 

This paper is organized as follows. In Section~\ref{sec:cascade} we 
summarize the calculation used in order to obtain the shower
development. The results are given in Section~\ref{sec:results}.
Finally, in Section~\ref{sec:discussion} the results are 
discussed\footnote{Throughout this paper we use natural units, \IE
$\hbar = c =1$}. A preliminary account of these results was given
in ref.~\cite{giant}
\section{The Cascade Equations}
\label{sec:cascade}
We use one dimensional cascade theory in Approximation A; due to the
fact that we are mainly interested in ultrahigh energies, this
should be adequate in order to obtain the first estimate of the
behavior of the showers induced by neutrino interactions.

The incoming neutrino interacts with a quark in the target nucleus
and generates a hadronic jet. In the present work, we assume that
the  hadrons generated share the primary energy eqally; hence we start with
showers induced by a single hadron. 

 In a hadronic interaction, most of the secondaries produced
are pions. Fast nucleons are relatively rare, with the possible exception of
the leading nucleon in a nucleon--nucleus interaction. At this point, 
we neglect the production of particles other than pions.
 
We assume that pions of either charge are produced in 
equal numbers. Since the average multiplicity in an
interaction is large, this is a permissible simplification.
 We neglect photoproduction of pions: hence, neutral pions act as 
the source of the electromagnetic component
without the latter affecting the hadronic one. As a consequence, 
the equations
governing  the hadronic component are decoupled from the ones
governing the electromagnetic one.

 We assume that Feynman scaling is valid and parametrize
the single particle inclusive distributions by means of
a simple expression,
\BEQ
\frac{1}{\sigma} E \frac{d\sigma}{dE} = F(z) = K z^{\alpha}(1-z)^{\beta}
\Theta \left( z - z_{0} \right).
\label{eq:inclusive}
\EEQ
Here, $z=E/E'$ is the Feynman parameter, $E'$ being the primary energy.
The quantities $K, \alpha, \beta $ are to be determined from the
normalization conditions,
\BEQ \int dE  \frac{d\sigma}{dE} = <n> \sigma_{inel}
\label{eq:multiplicity}
\EEQ
and
\BEQ \int  dE E  \frac{d\sigma}{dE} = E', 
\label{eq:energy}
\EEQ
and from fits to the experimental data. An infrared (IR) cutoff,
$ z_{0}$ has been introduced. Equation~(\ref{eq:multiplicity})
follows from the definition of a single particle inclusive cross section,
whereas (\ref{eq:energy}) expresses energy conservation.

Due to the fact that we treat all light hadrons together, some compromises 
are
necessary: in fact, the parameter $\beta$ governs the leading particle 
behavior which is different for the various secondaries. Nevertheless, the 
available data \cite{PDG} suggest that taking 
$\alpha = 0$ and  $\beta = 5$ gives a reasonable fit to all data. 
(By choosing $\alpha = 0 $ one gets an inclusive distribution
$ \propto 1/z$ for small values of the Feynman parameter. This 
is consistent with the fact that at low momenta the inclusive
distribution is dominated by soft gluon emission.) The  
infrared cutoff has been chosen at a fixed Feynman parameter rather than at a
fixed energy. This is consistent with the fact that in the following, we use an
energy independent average multiplicity, $<n> = 50$ --- again a compromise.
By doing so, one underestimates the number of soft secondaries. However,
we found that these approximations have virtually no effect on the
integral distibutions with a threshold energy $E_{th}\gappr 100$GeV.
Likewise, the shower development is rather insensitive to the value of
$\beta$: even a distribution $F(z) \propto \Theta (1 - z) $ gives 
reasonable showers. Likewise, the integral distributions at the threshold
energies indicated above are quite insensitive to the choice of 
the IR cutoff.
For this reason, we put $z_{0} = 0$ whenever this choice does not give
rise to IR divergences.

We are now ready to write down the transport equations for the hadronic
component of the shower. There are two distinct equations. The first one is
for the stable hadrons: nucleons and charged pions; at the relevant energies
the charged pions may be considered stable. The second equation characterizes
the behavior of neutral pions: in that case, decay is essential\footnote{One 
recalls that in air at ground level, the decay and interaction mean 
free paths of a neutral pion become equal at an energy 
of about $10^{18}$eV.}. We denote the differential distribution of
charged hadrons and neutral pions by $h$ and $p$, respectively. The
depth of the medium is measured in units of the hadronic interaction 
mean free path, $\tau = t/\lambda $, where $t$ is the target
thickness measured in ${\rm g/cm}^{2}$. The approximate transport
equations read as follows.
\begin{eqnarray}
\frac{\partial h}{\partial \tau} = & -h +2/3 \int \frac{dE'}{E} F\left(
\frac{E}{E'} \right) h\left( E'\right) \nonumber \\
\frac{\partial p}{\partial \tau} = & - \frac{ \lambda }{D  \rho}p
 + 1/3 \int \frac{d E'}{E} F\left( \frac{E}{E'}\right) h\left( E'\right).
\label{eq:hadrondiff} 
\end{eqnarray}

In writing down  equation~(\ref{eq:hadrondiff}) we neglected the 
interaction of neutral
pions. The loss term contains merely their decay. The factor 
$ \frac{\lambda}{\rho}$ serves to convert the real space decay length, D,
to the units used in writing down the transport equation. The decay length
is, of course, given by the expression:
\[ D = \frac{m_{0}}{T E}, \]
where $T$ is the lifetime in the rest frame and $m_{0}$ is the mass of the 
neutral pion.
The density
of the medium, $\rho$, is expressed as a function of $\tau$. The
mass of the neutral pion is denoted by $ m_{0} $. The factors 2/3 and 1/3
reflect  the fact that approximately two out of three  hadrons produced
are charged pions and one out of three is a neutral pion.

In a similar fashion, we can write down the equation governing the
electromagnetic component. To the accuracy desired for the purposes
of the present calculation, the (common) inclusive distribution
of Bremsstrahlung and pair production can be approximated by a
step function, $F(E/E')= \Theta (1 - E/E')$. (This gives, of course,
a pure $1/E$ inclusive spectrum for both the charged leptons and 
the Bremsstrahlung photons, a reasonably accurate approximation.)
In the same approximation, we put the cross sections of Bremsstrahlung
and pair production equal to each other (\IE $7/9 \approx 1$). We denote the
number of particles interacting only electromagnetically by $l$, \IE
\[ n_{e^{+}} + n_{e^{-}} + n_{\gamma} \approx l. \]
The latter approximate equality is valid after a few radiation lengths;
initially, the number of charged leptons is slightly overestimated.

We write down the the evolution equation for $l$ by measuring
distances in units of the radiation length, $X_{0}$.  Assuming again 
that all neutral pions decay instantaneously and that the energy of the
$\pi^{0}$ is shared equally between the decay photons, we have with
$ \sigma = t/X_{0}$:
\BEQ
\frac{\partial l(E, \sigma)}{\partial \sigma} = \frac{X_{0}}{D} 
p\left(2E, \frac{X_{0} \sigma}{\lambda}\right)
+ \int \frac{dE'}{E} \Theta (E' - E) l\left( E', \sigma\right) - 
l\left(E,\sigma\right)
\label{eq:leptondiff}
\EEQ
The evolution equations given by (\ref{eq:hadrondiff})
and (\ref{eq:leptondiff}) form a hierarchical structure. Due to the 
(very reasonable) approximation made in writing down these equations, 
namely that {\em neutral pions decay instantly and do not interact},
one has to solve only the first one of equations~(\ref{eq:hadrondiff});
the evolution equation for neutral pions can be solved by quadrature
and equation~(\ref{eq:leptondiff}) is solved in terms of a Green function.

In fact, the evolution equation for neutral pions is solved by the
expression:
\BEQ
p = \int d\tau ' {\rm e}^{- \left( \tau - \tau '\right)\lambda /D }
\Theta (\tau - \tau ') g(\tau'),
\label{eq:neutralpi}
\EEQ
where $g(\tau)$ is the gain term as one can read it off from the second of 
equations~(\ref{eq:hadrondiff}).
Due to the fact that $\lambda/D \gg 1$, one can pull out $g(\tau ')$
from under the integral in eq.~(\ref{eq:neutralpi}) at $\tau' = \tau$.
Upon substituting this into the evolution equation of leptons, 
eq.~(\ref{eq:leptondiff}), one realizes that in the rapid decay approximation,
that equation also scales as the hadronic evolution equations are 
supposed to do. Hence we conclude that  Feynman scaling holds for a 
mixed hadron - lepton
cascade if:
\begin{itemize}
\item The energies are sufficiently high so that effects of the rest masses
of the particles are negligibly small,
\item the energies are sufficiently low so that the reinteraction of
decaying neutral particles is negligible together with the QCD loop
corrections containing logarithms of the energy. (If the decay length
does not cancel, there is a scaling violation due to the Lorentz factor
present in the expression of the decay length.)
\end{itemize}

In order to find the retarded Green function of eq.~(\ref{eq:leptondiff}),
one uses Mellin transform techniques. The resultis: 
\begin{eqnarray}
G\left( q, \tau - \tau '\right) = \exp \left( \tau - \tau '\right)
\Theta \left( \tau - \tau '\right) \Theta \left( 1 - q\right)
\left( \frac{\ln |q|}{2 (\tau - \tau ')}\right)^{- 1/2} \nonumber \\
 \times I_{1}\left( 2^{3/2} \left(\tau - \tau '\right)^{1/2} (\ln |q|)^{1/2}
\right).
\label{eq:Green}
\end{eqnarray}
In eq.~(\ref{eq:Green}), $q$ stands for the scaling variable, 
$q = E/E'$ and $I$ is a  modified Bessel function of order one, \CF
\cite{Erdelyi}.

To summarize: assuming the validity
of Feynman scaling and the rapid decay of neutral pions, one  has to
solve one evolution equation only, \VIZ the one governing the
evolution of charged hadrons. Other components of the mixed
hadronic and leptonic cascade are obtained by quadratures, which is
rapidly accomplished by using standard numerical integration techniques.

The solution of the first of the equations (\ref{eq:hadrondiff}) is best
accomplished by iteration. In fact, one notices that by converting that 
equation into a Volterra integral equation, an iterative procedure 
rapidly converges. Roughly speaking, the number of iterations needed 
is about the number of collision mean free paths at which the distribution
is to be computed. (This has been known for a very long time:
Bhaba's method of ``succesive collisions'' for obtaining  the
evolution of  an electromagnetic 
cascade is, in essence, based on this observation, see, \EG~\cite{Rossi}.)
Equations for the integral distributions can be derived in a straight
forward manner; the calculation is particularly easy in the Feynman scaling
limit. 
\section{Results}
\label{sec:results}
We calculated the integral distributions of electrons and positrons for 
a number of hadron energies and nadir angles. In each case, we assumed that
the incident neutrino interacts in the Earth, just below its exit point.
(In practical terms, this means that the interaction has to take place 
within a few hadronic interaction mfp below the surface.) For each nadir 
angle,
the  target thickness penetrated  by a  neutrino before 
interaction was determined on the basis 
of the results given in ref.~\cite{gandhi}. The profile of the 
electromagnetic  component of the shower is shown in the following
Figures. In each calculation, the integral spectrum of the hadrons was
cut off at $E_{0} =100$GeV, corresponding to a CMS energy of
$\sqrt{s}\approx 14$GeV. Below such energies, particle production 
becomes insignificant; moreover, the particles produced tend to decay 
into soft muons and gammas instead of contributing to the cascade.
Likewise, the integral distribution of the electromagnetic component
was cut off at $E_{c}=100$MeV, roughly corresponding to the critical
energy in air. In estimating the primary neutrino energies, we
assumed an average inelasticity $<y> = 0.5$ and an average
hadronic multiplicity, $<n>=50$ in the first interaction. Thus, on the
average, $E_{hadron}=E_{\nu}/100$. The variation of the nadir angle around
the horizontal direction corresponds to a slight variation of the
target thickness around one neutrino mfp.

The following figures display the electron--positron component 
generated by a single hadron. Due to the linearity of the evolution
equations, the curves displayed here can be easily scaled to accommodate
other inelasticities and/or neutrino primary energies. The shower profile
is presented in ``real space''. For purposes of converting target depth 
to length an exponential atmosphere was used, with scale
height, $h=16.8$km and ground level density, 
$\rho_{0}= 1.3\times 10^{-3}{\rm g/cm^{3}}$.
The notation 
used in the Figures is summarized in Table~1.\vspace{5mm}
\begin{center}
\begin{tabular}{||l|c|c|c||} \hline
$E_{h}$ (GeV) &
\multicolumn{3}{c||}{Nadir angles}\\
\hline
 & diamonds & triangles  & asterisks\\ \hline
$10^{7}$  & \DEG{80}& \DEG{85}& \DEG{88}\\ \hline
$10^{8}$  & \DEG{80}& \DEG{84}& \DEG{87}\\ \hline
$10^{9}$ & \DEG{80} & \DEG{84}& \DEG{88}\\ \hline
\end{tabular}\\[3mm]
Table 1. Summary of hadron energies and symbols used for
showers developing at different nadir angles. The average single
hadron energy, $E_{h}$  is assumed to be 1\% of the incident
neutrino energy, see text.
\end{center}

\centerfigure{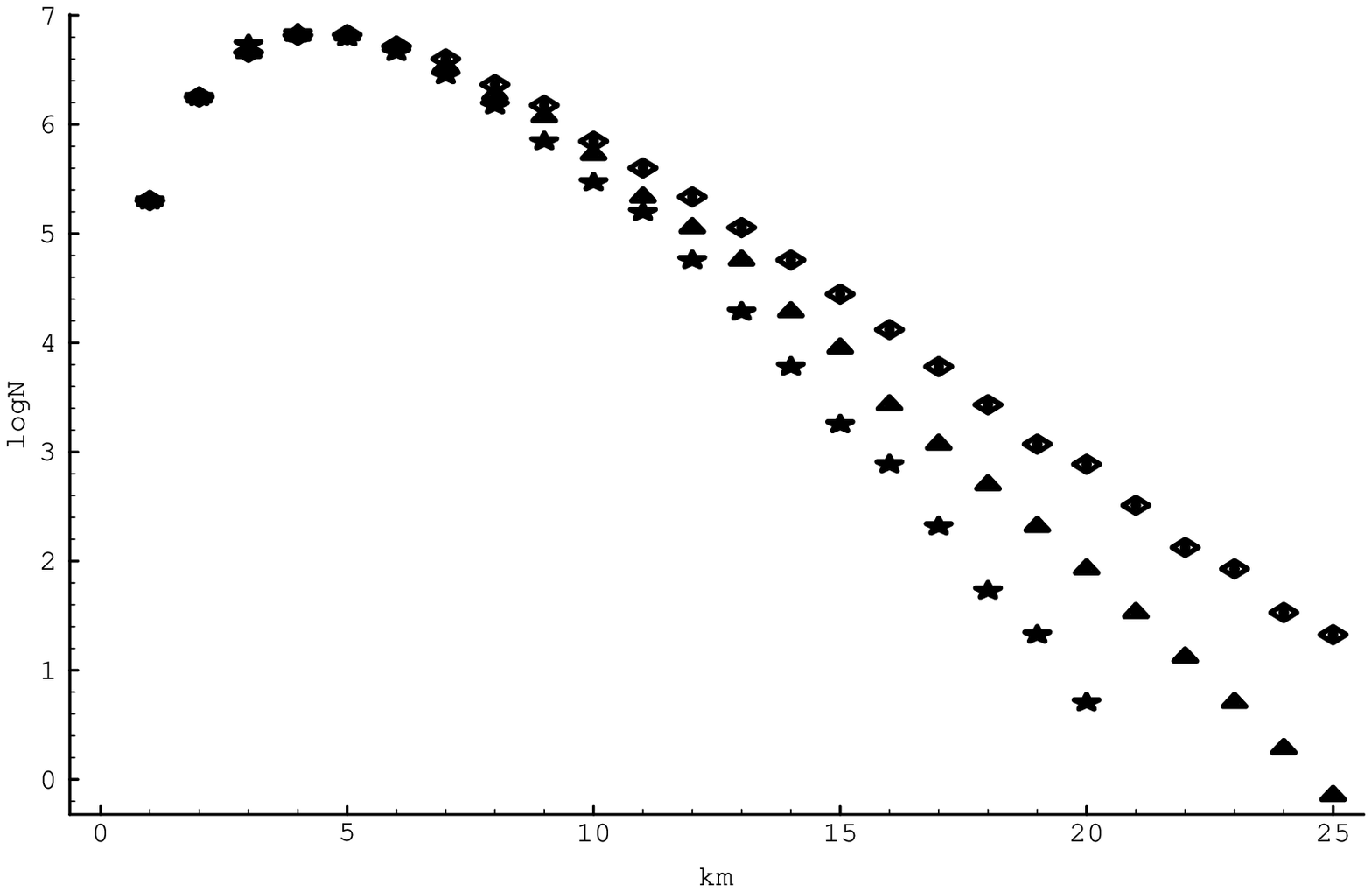}{Fig.1: Integral distribution of electrons
and positrons, $E>100$MeV. $E_{h}=10^{7}$GeV }
\vspace{3mm}
\centerfigure{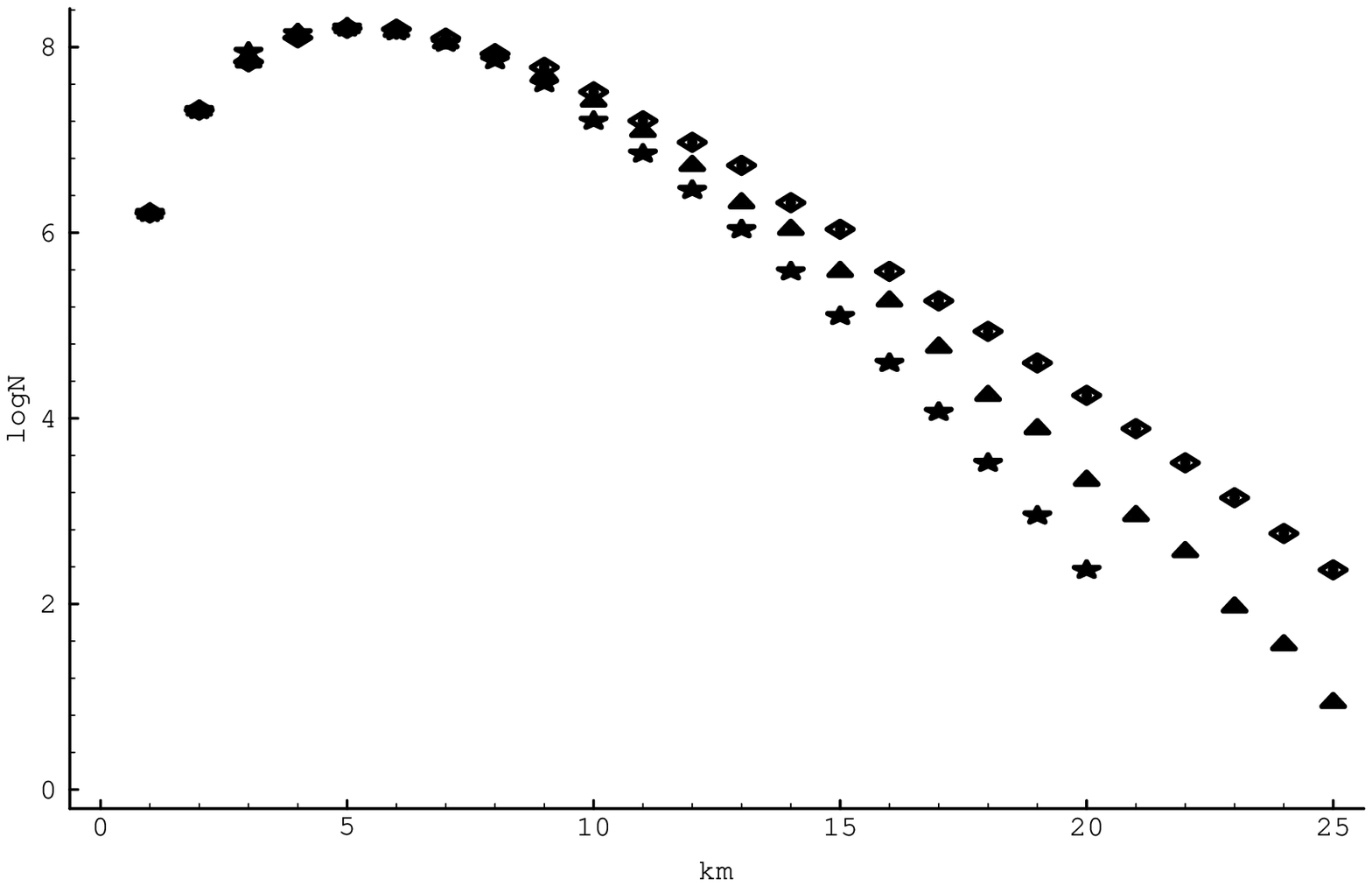}{Fig.2: Same, $E_{h}=10^{8}$GeV}
\vspace{3mm}
\centerfigure{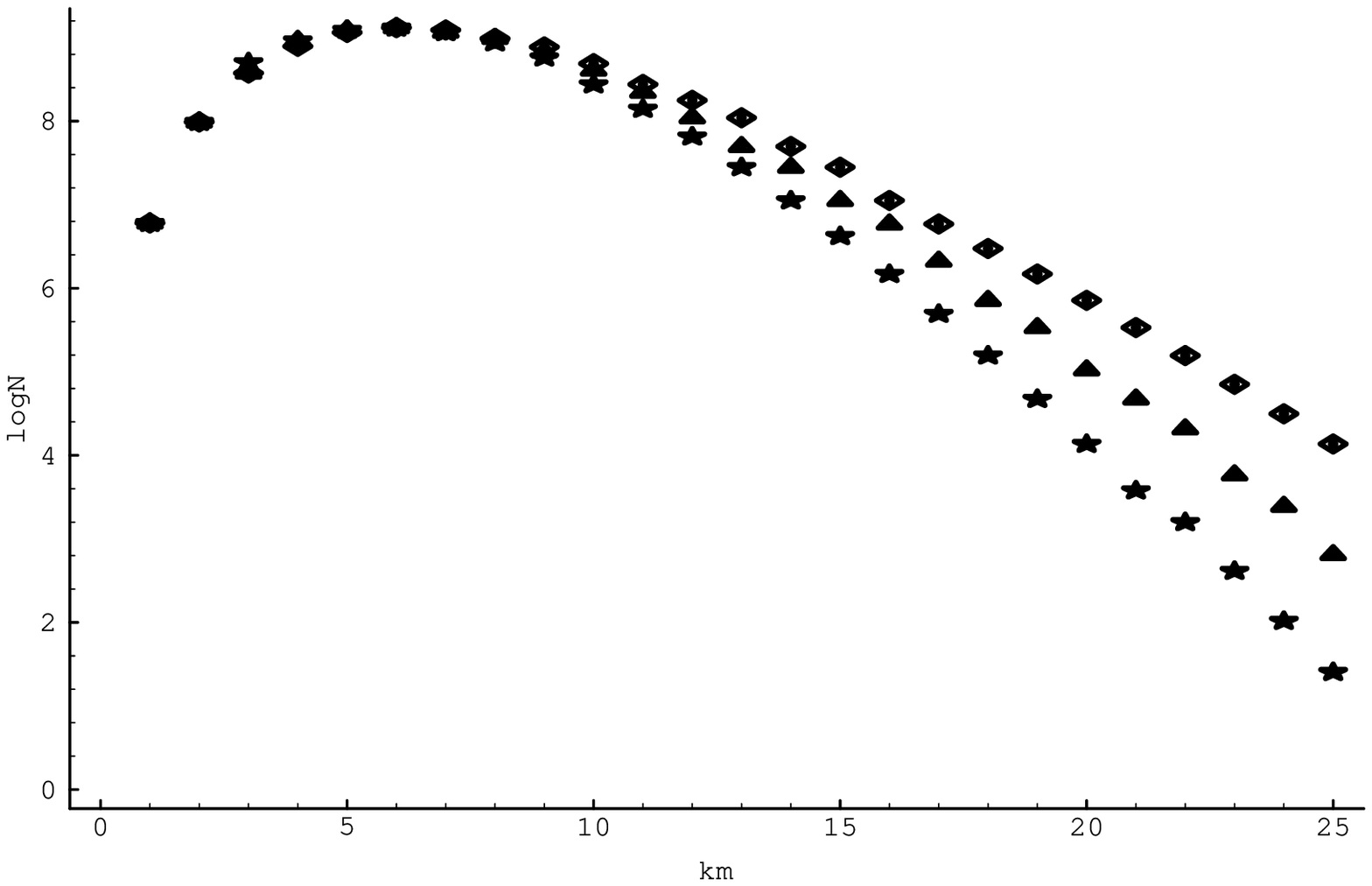}{Fig.3: Same, $E_{h}=10^{9}$GeV}
\vspace{5mm}
\section{Discussion}
\label{sec:discussion}
The figures displayed in the preceding section reveal a characteristic 
feature of the showers induced by a netrino interaction near the exit point
from the Earth. Those showers are ``upside--down'' -- they start in
the dense portion of the atmosphere and evolve upwards. As a consequence,
such showers reach their maximum rather soon, typically at distances
of the order of 5km. Thereafter, they have an unusually long tail, 
\IE a very slowly decreasing particle number as the shower evolves.
This is an effect which is easily understood: the ``upside--down''
shower starts near the surface of the Earth and proceeds upwards.
As far as the hadronic (charged pions and nucleons) and the purely
leptonic ($\gamma , e^{\pm}$) components are concerned, the patterns of a
``normal'' or of an ``upside--down''
shower are the same if plotted against the target depth. However, the
distribution of neutral pions (the main source of photons responsible for 
initiating the leptonic component) is asymmetric: upward and
downward going  beams of $\pi^{0}$-s can be easily distinguished
from each other. As a consequence, any orbiting detector which is able to
follow the longitudinal development of a shower is, {\em a fortiori}
able to pick up upward going showers.

It remains to be seen whether sufficiently powerful 
extraterrestrial neutrino sources exist so that one can collect a
significant number of neutrino events of the type described here.
Besides the obvious astrophysical interest of such events, one 
would be able to look for signatures of ``new physics'', assuming
that it affects the longitudinal development of the neutrino
induced showers. For instance, it was proposed that at some
higher energy scale, neutrinos begin to interact more strongly than
it would be expected on the basis of the standard electroweak theory
see refs.~\cite{domonuss, domo}\footnote{It was pointed out that, taken
literally, such schemes violate unitarity, see~\cite{burdman}. 
This is due to the fact that in refs.~\cite{domonuss, domo} the onset 
of ``new physics'' was modelled by adding a piece to the cross section
which is proportional to a step function. As a consequence, the real
part of the corresponding amplitude develops a logarithmic singularity.
Almost any smooth function approximating a  step function is free of
such a singularity and can be reconciled with unitarity.}. Should this be 
the case, the correlation between the impact parameter of the 
incident neutrino and the development of the shower would change.
Assuming a sufficiently large  statistical sample of ``anomalous'' events,
one could deduce the onset of a new physical phenomenon\footnote{Of
course, fluctuations can affect the results; this question is being studied.
However, at sufficiently high energies, the development of a shower
gives information on the energy of the shower on an event by event
basis as well, see, \EG~\cite{sokolsky}.}.

{\bf Acknowledgements} We benefitted from many conversations with
Bianca Monteleoni of the University of Florence, about problems 
and perspectives of neutrino detection.  Leon Madansky of this
University 
has been most helpful in discussing problems related to
the detection of  extraterrestrial
neutrino sources. Last but not least, we thank John~F. Krizmanic for
discussions regarding OWL.

\end{document}